\documentclass[onecolumn,showpacs,superscriptaddress,prd,showkeys]{revtex4}%
\usepackage{amsmath}
\usepackage{amsfonts}
\usepackage{amssymb}
\usepackage{graphicx}

\usepackage{color}

\def\b{\begin{equation}}
\def\e{\end{equation}}
\def\ba{\begin{eqnarray}}
\def\ea{\end{eqnarray}}

\def\gabf{\mbox{\boldmath{$\gamma$}}}

\def\u{\uparrow}
\def\d{\downarrow}
\def\le{\langle}
\def\re{\rangle}

\begin{document}
-

\title{Spin-Particles Entanglement in Robertson-Walker Spacetime}

\author{Shahpoor Moradi}
\affiliation{Department of Physics, Faculty of
Science, Razi University, Kermanshah, Iran}
\email{shahpoor.moradi@gmail.com}

\author{Roberto Pierini}
\affiliation{School of Science and Technology, University of Camerino, 62032 Camerino, Italy}
\affiliation{INFN Sezione di Perugia, 06123 Perugia, Italy}
\email{roberto.pierini@unicam.it}

\author{Stefano Mancini}
\affiliation{School of Science and Technology, University of Camerino, 62032 Camerino, Italy}
\affiliation{INFN Sezione di Perugia, 06123 Perugia, Italy}
\email{stefano.mancini@unicam.it}

\begin{abstract}
We study the entanglement between two modes of  Dirac field
in an expanding spacetime characterized by the Robertson-Walker metric.
This spacetime model turns out to be asymptotically (in the remote past and far future regions) Minkowskian.
Then, on the one hand we show entanglement creation between particles and anti-particles
when passing from remote past to far future. On the other hand we show that particles entanglement in the remote past degrades into the far future. These effects are traced back to particles creation.
In our analysis we highlight the role of spin (polarization) of particles and compare the results with those obtainable without accounting for it.
\end{abstract}

\pacs{04.62.+v, 03.67.Mn}
\keywords{Quantum fields in curved spacetime, Entanglement characterization}

\date{\today}

\maketitle

\section{Introduction}

Entanglement was recognized since the early days of quantum mechanics as ``the characteristic trait of quantum mechanics" \cite{Sch}.
In fact it is a property unique to quantum systems. Actually two systems (particles)
are said to be entangled if they are described by
a density operator that cannot be written as a weighted sum of product density operators.
In such a case, the two systems can be said not to have a state of their own, even though they may
be arbitrarily far apart. As a consequence they posses correlations that
go beyond what is classically possible.

Recently the notion of entanglement has been
re-discovered in the context of quantum information theory where it has been recognized as
a resource for quantum information processing \cite{Horo}.
This renewed interest on entanglement was mainly confined on non-relativistic scenarios.
Nevertheless, boosts for relativistic extensions of quantum information theory come from
 scenarios like black hole physics, quantum gravity and quantum cosmology \cite{PT04}.
Then, relativistic properties of entanglement have started to
be investigated within special relativity.
For instance, entanglement
was found to be an observer-dependent property that is
degraded from the perspective of moving observers \cite{JSS07,CCM12}.
These results suggest that also in curved spacetime entanglement
might not be an invariant \cite{FS05,AL06,L09,M09}.
There, the study of entanglement becomes particularly difficult.
In fact, differently from Minkowski spacetime, in curved spacetime one does
not have the Poincare symmetry group \cite{bdv}. This makes the
classification of particles rather ambiguous. Their definition
requires that in the distant past and in the far
future spacetime  tends to Minkowskian spacetime.
Taking this approach it has been possible to learn about certain aspects
of entanglement in curved spacetimes \cite{TU04,Shi04,ball,st}.
A suitable metric for a curved spacetime satisfying the above requirements turns out to be the (conformally flat)
Robertson-Walker metric \cite{bdv}. Then in Ref.\cite{iv} the creation of entanglement between Dirac modes due to the expansion of a Robertson-Walker spacetime has been studied. However, the analysis was confined to spin-less particles.
Here, resorting to the same spacetime model, we account for the spin degrees of freedom in studying entanglement properties of Dirac modes. Besides revisiting the possibility of entanglement creation, we also address the issue of entanglement degradation. All these effects are traced back to particles creation in the expanding spacetime.
Actually we evaluate, when passing from remote past to far future,
the amount of entanglement created between particles and anti-particles, as well as the amount of destroyed entanglement between particles. In doing that a comparison of results with the case of spin-less particles is provided.

The paper is organized as follows.
In Section \ref{expand} we introduce the Dirac spinors in Robertson-Walker spacetime.
Then Bogolyubov transformations between Minkowskian regions are derived in Section \ref{Bogo} where the effect of particles creation is also discussed. In Section \ref{creation} we address the issue of particles-anti-particles entanglement creation, while in
Section \ref{entaD} we discuss the degradation of entanglement.
Conclusions are drawn in Section \ref{conclu}.


\section{Dirac spinors on Robertson-Walker spacetime}\label{expand}

In curved spacetime the definition of particle
requires that in the distant past and in the far
 future spacetime  tends to Minkowskian spacetime \cite{bdv}. This happens if a  gravitational
field is present only for a certain period of time. In this case,
there are natural $in$ (remote past) and $out$ (far  future) vacuum
states. In these asymptotic $in$- and $out$-regions the influence of
curved space-time diminishes in such a way that the concept of a
particle may be introduced. The relation between the corresponding
mode functions is described by a certain set of Bogolyubov
coefficients. Since the choice of the $in$ and $out$ vacuum states
is unique for all inertial particle detectors, one obtain a total
number density of particles. This is the density of particles
produced by gravity in a spacetime where the field was initially  in
the natural vacuum state. The created particles are observed at late
times, when gravity is inactive and the definition of particles is
again possible.

To implement this procedure we start by considering the Robertson-Walker line element \cite{bdv}
\b
ds^2=a^2(\eta)(-d\eta^2+dx^2+dy^2+dz^2),
\label{ds}
\e
where the dimensionless
conformal time $\eta$ is related to cosmological time $t$ by
$\eta=\int{a^{-1}(t)dt}$. Here $a(\eta)$ is the scale factor determining the spacetime expansion rate. We assume flat spacetime as $\eta\rightarrow \pm\infty$.
The covariant generalization of Dirac equation  for Dirac field $\psi$ of mass $m$
on a curved background is given by
\b
\left(\tilde{\gamma}^{\mu}(\partial_{\mu}+\Gamma_{\mu})
+m\right)\psi=0,
\label{Deq}
\e
where the curved gamma matrices
$\tilde{\gamma}^{\mu}$ are related to flat ones through
$\tilde{\gamma}^{\mu}:=a^{-1}\gamma^{\mu}$ and the spin connections read
\b
\Gamma_{\mu}=\frac{1}{4}\frac{\dot{a}}{a}[\gamma_{\mu},\gamma_0].
\label{Gm}
\e
Here and below dot  denotes the derivative with respect to conformal time $\eta$.

Writing $\psi =a^{-3/2}(\gamma^{\nu}\partial_{\nu}-M)\varphi$, with $M=ma(\eta)$,
in \eqref{Deq} we get
\b
g^{\mu\nu}\partial_{\mu}\partial_{\nu}\varphi-
\gamma^{0}\dot{M}\varphi-M^2\varphi=0,
\label{newDeq}
\e
being $g^{\mu\nu}$ the flat metric as opposed to the actual spacetime metric $\tilde{g}^{\mu\nu}$.
Moreover, given the flat spinors $u_d$ and $v_d$ (with $d=\u,\d$)
satisfying the relations
\b
\gamma^0u_d=-iu_d, \quad\quad \gamma^0v_d=iv_d,
\label{gammarel}
\e
we set
\b
\varphi:=N^{(\mp)}f^{(\mp)}(\eta)u_d
e^{\mp i{\mathbf{p}}\cdot{\mathbf{x}}}, \quad {\rm or} \quad
\varphi:=N^{(\mp)}f^{(\mp)}(\eta)
v_d e^{\mp i{\mathbf{p}}\cdot{\mathbf{x}}},
\label{varphidef}
\e
with ${\mathbf{x}}$, ${\mathbf{p}}$ position and momentum vectors in $\mathbb{R}^3$.
The functions $f^{(\pm)}$ obey the differential equation
\b
\ddot{f}^{(\pm)}+
\left(|{\mathbf{p}}|^{2}+M^2\pm
i\dot{M}\right)f^{(\pm)}=0.
\label{feq}
\e
Actually  $f^{(-)}$ and $f^{(+)}$ are
positive and negative  frequency modes with respect to conformal
time $\eta$ near the asymptotic past and future, i.e.
$i\dot{f}^{(\pm)}(\eta)\approx \mp E_{in/out} f^{(\pm)}(\eta) $ with
\b
E_{in/out}=\sqrt{|{\mathbf{p}}|^2+M^2_{in/out}}.
\label{Einout}
\e
Now we can introduce the spinors that behave respectively like positive and
negative energy spinors in the asymptotic regions, namely

\begin{align}
U({\mathbf{p}},d,{\mathbf{x}},\eta)&:=U({\mathbf{p}},d,\eta)e^{-i{\mathbf{p}}\cdot{\mathbf{x}}}=N^{(-)}(\gamma^{\nu}\partial_{\nu}-M)f^{(-)}(\eta)u_de^{-i{\mathbf{p}}\cdot{\mathbf{x}}},
\label{Updeta}
\\
V({\mathbf{p}},d,{\mathbf{x}},\eta)&:=V({\mathbf{p}},d,\eta)e^{i{\mathbf{p}}\cdot{\mathbf{x}}}=N^{(+)}(\gamma^{\nu}\partial_{\nu}-M)f^{(+)}(\eta)v_de^{i{\mathbf{p}}\cdot{\mathbf{x}}},
\label{Vpdeta}
\end{align}
By defining  $\bar{U}:=iU^{\dag}\gamma^0$ and $\bar{V}:=iU^{\dag}\gamma^0$, we get
the orthonormality relations  $ \bar{U}_dU_{d'}=\delta_{dd'}, \bar{V}_dV_{d'}=-\delta_{dd'}$ that
lead to the normalization constants
\b
N^{(+)}=N^{(-)}=\frac{1}{\sqrt{2M_{in/out}(E_{in/out}+M_{in/out})}}.
\label{Npm}
\e
Then, the normalized spinors read
\begin{align}
U_{in/out}({\mathbf{p}},d,\eta)&=\frac{1}{\sqrt{2M_{in/out}(E_{in/out}+M_{in/out})}}
\left(-i\dot{f}^{(-)}(\eta)-Mf^{(-)}(\eta)+
if^{(-)}(\eta) \gabf \cdot {\mathbf{p}}\right)u_d,
\label{Uspinor}\\
V_{in/out}({\mathbf{p}},d,\eta)&=\frac{1}{\sqrt{2M_{in/out}(E_{in/out}+M_{in/out})}}\left(+i\dot{f}^{(+)}(\eta)
-Mf^{(+)}(\eta)-
if^{(+)}(\eta)  \gabf \cdot {\mathbf{p}}\right)v_d,
\label{Vspinor}
\end{align}
and in the asymptotic regions they reduce to the flat ones
\begin{align}
u_{in/out}({\mathbf{p}},d)&:=
\frac{i{\not p}-M_{in/out}}{\sqrt{2M_{in/out}(E_{in/out}+M_{in/out})}}u_d,
\label{uflat}\\
v_{in/out}({\mathbf{p}},d)&:= -\frac{i{\not
p}+M_{in/out}}{\sqrt{2M_{in/out}(E_{in/out}+M_{in/out})}}v_d.
\label{vflat}
\end{align}


\section{Bogolyubov transformation and particle creation}\label{Bogo}

Let us assume $\{U_{in},V_{in}\}$ and $\{U_{out},V_{out}\}$ to be two
complete sets of mode solutions for the Dirac equation \eqref{newDeq} which define
particles and anti-particles in asymptotic regions and have corresponding
vacua, $|0\re_{in}$ and $|0\re_{out}$ respectively.
 Physically, $|0\re_{in}$
is the state with no incoming particles (anti-particles) in remote
past and $|0\re_{out}$ is the state with no outgoing
particles (anti-particles) in the far future. When $\eta\rightarrow
\pm\infty$, the spacetime is flat and
 the dynamics of the field is that of the free field. So we have two natural
 quantization of the field, associated with two Fock spaces.
 The Dirac field operator can hence be written as
\ba
\psi({\mathbf{x}},\eta)&=& \int d{\mathbf{p}}\sum_{d}\left[a_{in}({\mathbf{p}},d)
U_{in}({\mathbf{p}},d,\eta)e^{i{\mathbf{p}}\cdot{\mathbf{x}}}+b^{\dag}_{in}({\mathbf{p}},d)V_{in}({\mathbf{p}},d,\eta)e^{-i{\mathbf{p}}\cdot{\mathbf{x}}}\right],
\label{psiin}\\
&=&\int d{\mathbf{p}}\sum_{d}\left[a_{out}({\mathbf{p}},d)
U_{out}({\mathbf{p}},d,\eta)e^{i{\mathbf{p}}\cdot{\mathbf{x}}}+b^{\dag}_{out}({\mathbf{p}},d)V_{out}({\mathbf{p}},d,\eta)
e^{-i{\mathbf{p}}\cdot{\mathbf{x}}}\right],
\label{psiout}
\ea
 where $a_{in},b_{in}$ and  $a_{out},b_{out}$
 are annihilation operators of particles and anti-particles in the $in$ and $out$ asymptotic
 regions respectively.
Actually $a_{in}$  and $b_{in}$  differ from
$a_{out}$  and $b_{out}$ because
 they do not correspond to physical particles outside the
 $in$ region. However, it is possible to relate the operators of
$in$-particles to those of $out$-particles by Bogolyubov transformation
\cite{bdv}. Since each set of spinors is
complete we can write down one set in terms of the other, e.g.
\begin{align}
U_{in}({\mathbf{p}},d,\eta)&={\mathcal{A}}({\mathbf{p}})U_{out}({\mathbf{p}},d,\eta)+\beta_{d,-d}({\mathbf{p}})V_{out}(-{\mathbf{p}},d,\eta),\label{Uinvsout}\\
V_{in}({\mathbf{p}},d,\eta)&=\rho_{d,-d}({\mathbf{p}})U_{out}(-{\mathbf{p}},d,\eta)+{\mathcal{C}}({\mathbf{p}})V_{out}({\mathbf{p}},d,\eta),\label{Vinvsout}
\end{align}
where $-d$ stands for the opposite spin projection of $d$.
These  relations and the complex coefficients ${\cal A}$, ${\cal C}$, $\beta_{dd'}$, $\rho_{dd'}$
 are called respectively \emph{Bogolyubov transformations} and \emph{Bogolyubov coefficients}. As consequence,
 the $in$ and $out$ ladder operators of particles and anti-particles are related by the unitary transformation
\b
\label{ainout}
\left(
  \begin{array}{c}
    a_{in}({\mathbf{p}},d) \\
    b^{\dag}_{in}(-{\mathbf{p}},-d) \\
  \end{array}
\right)=\left(
          \begin{array}{cc}
            {\mathcal{A}}^{\ast}({\bf p}) & \beta^{\ast}_{d,-d}({\bf p}) \\
            \rho^*_{-d,d}({-\bf p})  & {\mathcal{C}}^*(-{\bf p} ) \\
          \end{array}
        \right)\left(
                 \begin{array}{c}
                   a_{out}({\mathbf{p}},d) \\
                   b^{\dag}_{out}(-{\mathbf{p}},-d) \\
                 \end{array}
               \right).
\e
By using anticommutation relations we obtain that the entries of such unitary matrix (the Bogolyubov
coefficients) satisfy the following relations
\begin{align}
|{\mathcal{A}}({\mathbf{p}})|^2+|\beta_{d,-d}({\mathbf{p}})|^2&=1,
\label{albeort}\\
|\rho_{d,-d}({\mathbf{p}})|^2+|{\mathcal{C}}({\mathbf{p}})|^2&=1,
\label{rhosigort}\\
{\mathcal{A}}^{\ast}({\mathbf{p}})\rho_{d,-d}(-{\mathbf{p}})+
\beta^{\ast}_{d,-d}({\mathbf{p}}){\mathcal{C}}(-{\mathbf{p}})&=0.
\label{alrhoort}
\end{align}
In the $in$ region, the spacetime is  Minkowskian
and the vacuum of the field is $|0\re_{in}$, but this state
is not regarded by inertial observers in the $out$ region as the physical
vacuum (this role being reserved to the state $|0\re_{out}$).
We can therefore describe this
quantum `evolution' as creation of particles due to
spacetime expansion. The expectation value of the $out$
number of particles in the $in$ vacuum (i.e., number of created particles) with momentum
${\mathbf{p}}$ and spin projection $d$ is
\b
n^p({\mathbf{p}},d):=\le 0_{in}
|a^{\dag}_{out}({\mathbf{p}},d)a_{out}({\mathbf{p}},d)|0_{in}
\re =|\rho_{d,-d}(-{\mathbf{p}})|^2.
\label{np}
\e
Similarly, the
expectation value of $out$ number of anti-particles in the $in$ vacuum (i.e.,
number of created anti-particles) is
\b
n^a({\mathbf{p}},d):=\le 0_{in}|
b^{\dag}_{out}({\mathbf{p}},d)b_{out}({\mathbf{p}},d)|0_{in} \re=|\beta_{d,-d}(-{\mathbf{p}})|^2.
\label{na}
\e
Then, the total number of
created particles with momentum ${\mathbf{p}}$ (total number density) results
by combining \eqref{np} and \eqref{na} and summing up over spin projections
\b
\label{densityn}
n({\mathbf{p}}):=\sum_d\ \left[ n^p({\mathbf{p}},d)+n^a({\mathbf{p}},d)\right]
=\sum_d\ \left[ |\rho_{d,-d}(-{\mathbf{p}})|^2+|\beta_{d,-d}(-{\mathbf{p}})|^2\right].
\e

Being $f^{(-)}(\eta)$ and $f^{(+)}(\eta)$ positive and negative
frequency modes in asymptotic regions, we can write the Bogolyubov transformation
between them as follows
\ba
f^{(-)}_{in}(\eta)&=&A^{(-)}(p)f^{(-)}_{out}(\eta)+B^{(-)}(p)f^{(+)}_{out}(\eta),
\label{finm}\\
f^{(+)}_{in}(\eta)&=&A^{(+)}(p)f^{(+)}_{out}(\eta)+B^{(+)}(p)f^{(-)}_{out}(\eta).
\label{finp}
\ea
Since $f^{(+)}_{in}(\eta)=f^{\ast(-)}_{in}(\eta)$  we have
\b
A^{(+)}=A^{\ast(-)},\quad B^{(+)}=B^{\ast(-)}.
\label{Apm}
\e
By inserting \eqref{finm} and \eqref{finp} in \eqref{Uspinor} and \eqref{Vspinor} and using Bogolyubov transformations
\eqref{Uinvsout} and \eqref{Vinvsout} in limit $\eta\rightarrow +\infty $, we obtain  the
Bogolyubov coefficients as follows
\begin{align}
{\mathcal{A}}({\mathbf{p}})&={\mathcal{C}}^*({\mathbf{p}})=\sqrt{
\frac{E_{out}}{E_{in}}\frac{E_{out}+\mu_{out}}{E_{in}+\mu_{in}}}A^{(-)},
\label{alpsig}\\
\beta_{-d,d}({\mathbf{p}})&=-\rho_{d,-d}^*(-{\mathbf{p}})
=i\frac{v^{\dag}_{-d}\gamma\cdot{\mathbf{p}}u_d}
{\sqrt{\frac{E_{in}}{E_{out}}(E_{out}+\mu_{out})(E_{in}+\mu_{in})}}B^{(-)},
\label{betarho}
\end{align}
where $E_{in/out}$ are given by \eqref{Einout}.
As result \eqref{ainout} becomes
\b
\label{ainoutnew}
\left(
  \begin{array}{c}
    a_{in}({\mathbf{p}},d) \\
    b^{\dag}_{in}(-{\mathbf{p}},-d) \\
  \end{array}
\right)=\left(
          \begin{array}{cc}
            {\mathcal{A}}^{\ast} & \beta^{\ast}_{d,-d} \\
            -\beta_{d,-d} & {\mathcal{A}} \\
          \end{array}
        \right)\left(
                 \begin{array}{c}
                   a_{out}({\mathbf{p}},d) \\
                   b^{\dag}_{out}(-{\mathbf{p}},-d) \\
                 \end{array}
               \right).
\e


\section{Entanglement Creation}\label{creation}

Here we want to investigate the possibility of particle and anti-particle entanglement when passing from
$in$ to $out$ region. Then, to find the relation between $in$ and $out$ states we proceed as follows.
First, notice that the vacuum state in the $in$ region is a many particles state in the $out$ region due to particles creation during the expansion. Then if we restrict our attention to a single mode of momentum $p$
we can limit the occupation number to $2$ in the $out$ region, i.e. vacuum, one particle, and two particles
states can exist in the $out$ region. The same holds true for anti-particles of mode $-p$. Consequently
we would have
\b
\label{expansionABCD}
 |0_p;0_{-p}\re_{in}=A|0_p;0_{-p}\re_{out}+B|\u_p;\d_{-p}\re_{out}+C|\d_p;\u_{-p}\re_{out}+D|\u\d_p,\u\d_{-p}\re_{out},
 \e
 where, besides particle and antiparticle vacuum in the out region $|0_p;0_{-p}\re_{out}$, we have
 \begin{align}
 |\u_p;\d_{-p}\re_{out}&:=a^{\dag}_{out}({\mathbf{p}},\u)b^{\dag}_{out}(-{\mathbf{p}},\d)
 |0_p;0_{-p}\re_{out}
 =-b^{\dag}_{out}(-{\mathbf{p}},\d)a^{\dag}_{out}({\mathbf{p}},\u)|0_{-p};0_p\re_{out},\\
 |\u\d_p;0_{-p}\re_{out}&:=a^{\dag}_{out}({\mathbf{p}},\u)a^{\dag}_{out}({\mathbf{p}},\d)|0_p;0_{-p}\re_{out}=-
a^{\dag}_{out}({\mathbf{p}},\d)a^{\dag}_{out}({\mathbf{p}},\u)|0_{p};0_{-p}\re_{out}=-|\d\u_p;0_{-p}\re_{out},
 \end{align}
 and $A,B,C,D$ are coefficients to determine.

In writing Eq.\eqref{expansionABCD} we have used the fact that in the r.h.s. there must be no neat angular momentum
likewise the l.h.s. Then, the coefficients $A,B,C,D$ can be derived by noticing that $a_{in}({\mathbf{p}},d)$ acting on the vacuum gives zero and by using \eqref{ainoutnew}. It turns out that
\b
\label{0inout}
|0_p;0_{-p}\re_{in}=|{\mathcal{A}}|^2\left(|0_p;0_{-p}\re_{out}-\frac{\beta^*_{\u\d}}{{\mathcal{A}}^*}|\u_p;\d_{-p}\re_{out}
-\frac{\beta^*_{\d\u}}{{\mathcal{A}}^*}|\d;\u_{-p}\re_{out}+
\frac{\beta^*_{\u\d}\beta^*_{\d\u}}{{\mathcal{A}^*}^2}|\u\d_p,\u\d_{-p}\re_{out}\right),
\e
where we have omitted the explicit momentum dependence of Bogolyubov coefficients ${\cal A}$ and $\beta_{d,-d}$.

By means of \eqref{ainoutnew} and \eqref{0inout} we can also get
\begin{align}
\label{12inout}
|\u_p;0_{-p}\re_{in}:= a_{in}^{\dag}({\bf p},\u)|0_p;0_{-p}\re_{in}&=
\frac{\cal A}{{\cal A}^*}\left({\cal A}^* |\u_p;0_{-p}\re_{out}
-\beta^*_{\u\d} |\u\d_p; \u_{-p}\re_{out}\right),\nonumber\\
|\d_p;0_{-p}\re_{in}:= a_{in}^{\dag}({\bf p},\d)|0_p;0_{-p}\re_{in}&=
\frac{\cal A}{{\cal A}^*}\left({\cal A}^* |\d_p;0_{-p}\re_{out}
-\beta^*_{\u\d} |\u\d_p; \d_{-p}\re_{out}\right),\nonumber\\
|\u\d_p;0_{-p}\re_{in}:=  a_{in}^{\dag}({\bf p},\u)  a_{in}^{\dag}({\bf p},\d)|0_p;0_{-p}\re_{in}&=
\frac{\cal A}{{\cal A}^*} |\u\d_p;0_{-p}\re_{out},\nonumber\\
|\d\u_p;0_{-p}\re_{in}:=  a_{in}^{\dag}({\bf p},\d)  a_{in}^{\dag}({\bf p},\u)|0_p;0_{-p}\re_{in}&=
\frac{\cal A}{{\cal A}^*} |\d\u_p;0_{-p}\re_{out}.
\end{align}
The particle anti-particle density operator corresponding to the $out$ region state
\eqref{0inout} reads
\begin{align}
\label{rhooutcre}
\varrho^{(out)}_{p,-p}=\frac{1}{2}\Big[&
|{\mathcal{A}}|^4 |0_p;0_{-p}\re\le 0_p;0_{-p}|
-2{\mathcal{A}} {\mathcal{A}}^{* 2}\beta_{\u\d}
 |0_p;0_{-p}\re\le\u_p;\d_{-p}| \nonumber\\
& -2{\mathcal{A}} {\mathcal{A}}^{* 2}\beta_{\d\u}
 |0_p;0_{-p}\re\le\d_p;\u_{-p}|
  +2{\mathcal{A}}^{* 2} \beta_{\u\d} \beta_{\d\u}
 |0_p;0_{-p}\re\le \u\d_p;\u\d_{-p}|\nonumber\\
 &+|{\mathcal{A}}|^{2} |\beta_{\u\d}|^2
 |\u_p;\d_{-p}\re\le\u_p;\d_{-p}|
 +2|{\mathcal{A}}|^{2} \beta_{\u\d}^*\beta_{\d\u}
 |\u_p;\d_{-p}\re\le\d_p;\u_{-p}|
  -2{\mathcal{A}}^{*} |\beta_{\u\d}|^2 \beta_{\d\u}
 |\u_p;\d_{-p}\re\le\u\d_p;\u\d_{-p}|\nonumber\\
  &+|{\mathcal{A}}|^{2} |\beta_{\d\u}|^2
 |\d_p;\u_{-p}\re\le\d_p;\u_{-p}|
  -2{\mathcal{A}}^{*} \beta_{\u\d} |\beta_{\d\u}|^2
|\d_p;\u_{-p}\re\le\u\d_p;\u\d_{-p}|\nonumber\\
 &+ |\beta_{\u\d}|^2 |\beta_{\d\u}|^2
|\u\d_p;\u\d_{-p}\re\le\u\d_p;\u\d_{-p}| \Big] + {\rm h.c.}.
\end{align}
To evaluate the amount of entanglement of \eqref{rhooutcre} we use the \emph{logarithmic negativity} \cite{VW}.
It is defined as
\b
\label{deflogneg}
N(\varrho):=\log_2 \|\varrho^{PT}\|_1,
\e
where $\|\varrho^{PT}\|_1$ is the trace-norm of the partial (with respect to only one subsystem) transpose of $\varrho$, i.e. the sum of the absolute values of the eigenvalues of $\varrho^{PT}$ (since it is hermitian).

Notice that from \eqref{np} we have
\b
|\beta_{\u\d}|^2= n^p(\d), \quad |\beta_{\d\u}|^2= n^p(\u).
\e
Then, refereeing to \eqref{densityn} and assuming
$n^p(\u)=n^p(\d)=n^a(\u)=n^a(\d)=n/4$, we obtain the coefficients appearing in \eqref{rhooutcre} solely depending on $n$, that is
\b
\label{An}
|{\cal A}|^2=\frac{4-n}{4},\quad
|\beta_{\u\d}|^2 = |\beta_{\d\u}|^2 = \frac{n}{4}.
\e
Notice further that the density operator \eqref{rhooutcre}
lives in the Hilbert space $\mathbb{C}^4\times \mathbb{C}^4$.
Taking the basis $\{|0_p\re, |\u_p\re, |\d_p\re, |\u\d_p\re\}\otimes \{|0_{-p}\re, |\u_{-p}\re, |\d_{-p}\re, |\u\d_{-p}\re\}$
we can represent it  as a $16\times 16$ matrix and then compute the eigenvalues of its partially transpose
according to \eqref{deflogneg}.
Finally it results
\b
\label{LNspincreation}
L_N\left(\varrho^{(out)}_{p,-p}\right)=2\log_2\left[1+\frac{1}{2}\sqrt{n(4-n)}\right].
\e
It is worth remarking that $L_N$ is a concave function of $n$ taking minimum value (zero) for $n=0$ and $n=4$ and
maximum value (two) for $n=2$.


\subsection{Entanglement creation for spin-less particles}\label{entaC}

In the case of spinless particles we have two Bogolyubov coefficients
$\cal A$ and $\beta$ no longer depending on spin indexes and satisfying
$|{\cal A}|^2+|\beta|^2=1$
similarly to \eqref{albeort}. Moreover,
 refereeing again to \eqref{densityn} we now assume
$n^p=n^a=n/2$ and
Eq.\eqref{An} becomes
\b
\label{An2}
|{\cal A}|^2=\frac{2-n}{2},\quad
|\beta|^2 = \frac{n}{2}.
\e
To find the relation
between $in$ and $out$ states we proceed as above, writing
\b
|0_p;0_{-p}\re_{in}=A|0_p;0_{-p}\re_{out}+B|1_p;1_{-p}\re_{out},
\label{invacCC}
\e
with unknown coefficients $A, B$.
In this case Bogolyubov transformation \eqref{ainoutnew} reads
\b
\left(\begin{array}{c}
a_{in}(p )\\
b^{\dag}_{in}(-p)
\end{array}\right)
=
\left(\begin{array}{cc}
{\cal A}^{\ast} & \beta^{\ast} \\
\beta & {\cal A}
\end{array}\right)
\left(\begin{array}{c}
a_{out}(p )\\
b^{\dag}_{out}(-p)
\end{array}
\right).
\label{ainoutspinless}
\e
Then, imposing $a_{in}(p)|0_p;0_{-p}\re_{in}=0$ and using Eq.\eqref{ainoutspinless}
we can determine $A$ and $B$ obtaining
\ba
|0_p;0_{-p}\re_{in}&=&\frac{1}{\sqrt{1+|\beta/{\cal A}|^2}}(|0_p;0_{-p}\re_{out}-\frac{\beta^*}{\cal A^*}|1_p;1_{-p}\re_{out}).
\label{invac}
\ea
It results further
\b
|1_p;0_{-p}\re_{in}=|1_p;0_{-p}\re_{out}.
\label{in1p}
\e

The density operator in the $out$ region corresponding to \eqref{invac} reads
\b
\varrho^{(out)}_{p,-p}=\frac{1}{2}\Big[
\frac{|{\cal A}|^2}{|{\cal A}|^2+|\beta|^2} |0_p;0_{-p}\re \le 0_p;0_{-p}|
-2\frac{\beta {\cal A}^*}{|{\cal A}|^2+|\beta|^2}  |0_p;0_{-p}\re\le 1_p;1_{-p} |
+\frac{|\beta|^2}{|{\cal A}|^2+|\beta|^2}  |1_p;1_{-p}\re \le 1_p;1_{-p}|\Big] +{\rm h.c.},
 \label{rhoinvac}
\e
and it lives in the Hilbert space $\mathbb{C}^2\otimes\mathbb{C}^2$. Here, taking the basis $\{|0_p\re, |1_p\re\}\otimes \{|0_{-p}\re, |1_{-p}\re\}$
we can represent it  as a $4\times 4$ matrix and evaluate the eigenvalues of its partially transposed
according to \eqref{deflogneg}. Then, the amount of entanglement of \eqref{rhoinvac} results
\b
L_N\left(\varrho^{(out)}_{p,-p}\right)=\log_2\left[1+\sqrt{n(2-n)}\right].
\label{LNpa}
\e
Also in this case $L_N$ results a concave function of $n$ taking however minimum value (zero) for $n=0$ and $n=2$ and maximum value (one) for $n=1$.
 Notice that this result for spinless particles entanglement creation is consistent with the one that can be drawn in terms of subsystem's entropy from Ref.\cite{iv}.


\section{Entanglement Degradation}\label{entaD}

In this Section we want to study how
the entanglement between particles is affected by
 `evolution' from $in$ to $out$-region.

Let us start from two-mode maximally entangled state for particles and vacuum for antiparticles
\b\label{maxentpq}
|\Phi_{p,q;-p,-q}\re_{in} = \frac{1}{2}(|0_p,0_q\re + |\u_p,\u_q\re + |\d_p,\d_q\re + |\u\d_p,\u\d_q\re)_{in}\; |0_{-p},0_{-q}\re_{in}.
\e
For the sake of simplicity we assume Bogolyubov coefficients to be smooth varying functions of momentum so that choosing $p$ and $q$ to be close each other we have $\beta_{d,-d}(p )\approx\beta_{d,-d}(q)$ and ${\mathcal{A}}(p )\approx {\mathcal{A}}(q)$. Being these coefficients effectively dependent only on one argument,
we omit it in the following.
Then, using the in-out relations \eqref{0inout} and \eqref{12inout} on each mode, we get
\begin{align}
|\Phi_{p,q;-p,-q}\re_{in}
=\frac{|{\mathcal A}|^4}{2} \Big [ &
|0_p,0_{q}\re|0_{-p},0_{-q}\re
-\frac{\beta^*_{\u\d}}{{\mathcal A}^*}|0_p,\u_q\re|0_{-p},\d_{-q}\re
-\frac{\beta^*_{\d\u}}{{\mathcal A}^*}|0_p,\d_q\re|0_{-p},\u_{-q}\re
+ \frac{\beta^*_{\u\d}\beta^*_{\d\u}}{{{\mathcal A}^*}^2}|0_p,\u\d_q\re|0_{-p},\u\d_{-q}\re \nonumber\\
&- \frac{\beta^*_{\u\d}}{{\mathcal A}^*}|\u_p,0_q\re|\d_{-p},0_{-q}\re
+\frac{{{\beta^*}^2_{\d\u}}}{{{\mathcal A}^*}^2}|\u_p,\u_q\re|\d_{-p},\d_{-q}\re
+ \frac{\beta^*_{\u\d}\beta^*_{\d\u}}{{{\mathcal A}^*}^2}|\u_p,\d_q\re|\d_{-p},\u_{-q}\re \nonumber\\
&- \frac{{\beta^*}^2_{\u\d}\beta^*_{\d\u}}{{{\mathcal A}^*}^3}|\u_p,\u\d_q\re|\d_{-p},\u\d_{-q}\re
- \frac{{\beta^*}^2_{\d\u}}{{\mathcal A}^*}|\d_p,0_q\re|\u_{-p},0_{-q}\re
+ \frac{\beta^*_{\u\d}\beta^*_{\d\u}}{{{\mathcal A}^*}^2}|\d_p,\u_q\re|\u_{-p},\d_{-q}\re \nonumber\\
&- \frac{{\beta^*}^2_{\d\u}\beta^*_{\u\d}}{{{\mathcal A}^*}^3}|\u_p,\u\d_q\re|\d_{-p},\u\d_{-q}\re
+ \frac{{\beta^*}^2_{\d\u}}{{{\mathcal A}^*}^2}|\d_p,\d_q\re|\u_{-p},\u_{-q}\re
- \frac{\beta^*_{\u\d}{\beta^*}^2_{\d\u}}{{{\mathcal A}^*}^3}|\d_p,\u\d_q\re|\u_{-p},\u\d_{-q}\re \nonumber\\
&+ \frac{{\beta^*}_{\d\u}\beta^*_{\u\d}}{{{\mathcal A}^*}^2}|\u\d_p,0_q\re|\u\d_{-p},0_{-q}\re
+ \frac{{\beta^*}^2_{\d\u}}{{{\mathcal A}^*}^2}|\d_p,\d_q\re|\u_{-p},\u_{-q}\re
- \frac{\beta^*_{\d\u}{\beta^*}^2_{\u\d}}{{{\mathcal A}^*}^3}|\u\d_p,\u_q\re|\u\d_{-p},\d_{-q}\re \nonumber\\
&- \frac{{\beta^*}^2_{\d\u}\beta^*_{\u\d}}{{{\mathcal A}^*}^3}|\u\d_p,\d_q\re|\u\d_{-p},\u_{-q}\re
+ \frac{{{\beta^*}^2_{\u\d}}{\beta^*}^2_{\d\u}}{{{\mathcal A}^*}^2}|\u\d_p,\u\d_q\re|\u\d_{-p},\u\d_{-q}\re \Big ]_{out}  \nonumber\\
&\hspace{-1.2cm}+ \frac{{\mathcal A}}{2 {{\mathcal A}^*}^2} \Big[
{{\mathcal A}^*}^2|\u_p,\u_{q}\re|0_{-p},0_{-q}\re- \beta^*_{\d\u}{{\mathcal A}^*}|\u_p,\u\d_q\re|0_{-p},\u_{-q}\re - \beta^*_{\d\u}{{\mathcal A}^*}|\u\d_p,\u_q\re|\u_{-p},0_{-q}\re   \nonumber\\
&+ {\beta^*}^2_{\d\u}|\u\d_p,\u\d_q\re|\u_{-p},\u_{-q}\re +{{\mathcal A}^*}^2|\d_p,\d_q\re|0_{-p},0_{-q}\re + {\beta^*}_{\u\d}{\mathcal A}^*|\d_p,\u\d_q\re|0_{-p},\d_{-q}\re   \nonumber  \\
&+ \beta^*_{\u\d}{\mathcal A}^*|\u\d_p,\d_q\re|\d_{-p},0_{-q}\re
+ {\beta^*}^2_{\u\d}|\u\d_p,\u\d_q\re|\d_{-p},\d_{-q}\re+|\u\d_p,\u\d_q\re|0_{-p},0_{-q}\re \Big]_{out}.
\label{Phipq}
\end{align}
Then performing the trace over anti-particles (modes $-p,-q$) in the $out$ region
we get the particles state there
$\varrho^{(out)}_{p,q}={\rm Tr}_{-p-q} (|\Phi\re_{in}\le \Phi|)$.
Explicitly it is
\begin{align}
\varrho^{(out)}_{p,q} = \frac{1}{8}\Big[&
|{\mathcal A}|^8|0_p,0_q\re\le 0_p,0_q|
+2{{\mathcal A}^*}^4{{\mathcal A}}^2|0_p,0_q\re\le\u_p,\u_q|
+2{{\mathcal A}^*}^4{{\mathcal A}}^2|0_p,0_q\re\le\d_p,\d_q|
+2{{\mathcal A}^*}^4|0_p,0_q\re\le\u\d_p,\u\d_q| \nonumber  \\
&+|{\mathcal A}|^6|\beta_{\u\d}|^2 |0_p,\u_q\re\le0_p,\u_q|
-2{{\mathcal A}^*}^3{\mathcal A}|\beta_{\u\d}|^2|0_p,\u_q\re\le\d_p,\u\d_q|
\nonumber\\
&+|{\mathcal A}|^6|\beta_{\d\u}|^2|0_p,\d_q\re\le0_p,\d_q|
+2{{\mathcal A}^*}^3{\mathcal A}|\beta_{\d\u}|^2|0_p,\d_q\re\le\u_p,\u\d_q|
\nonumber\\
&+|{\mathcal A}|^4|\beta_{\d\u}|^2|\beta_{\u\d}|^2|0_p,\u\d_q\re\le0_p,\u\d_q|\nonumber\\
&+|{\mathcal A}|^6|\beta_{\u\d}|^2|\u_p, 0_q\re\le\u_p, 0_q|
-2{{\mathcal A}^*}^3{\mathcal A}|\beta_{\u\d}|^2|\u_p,0_q\re\le\u\d_p,\d_q| \nonumber \\
&+|{\mathcal A}|^4 \left(|\beta_{\u\d}|^4 +1\right) |\u_p,\u_q\re\le\u_p,\u_q|
+2|{\mathcal A}|^4 |\u_p,\u_q\re\le\d_p,\d_q|
+2{{\mathcal A}^*}^2 \left(|\beta_{\u\d}|^4+1\right) |\u_p,\u_q\re\le\u\d_p,\u\d_q|\nonumber  \\
&+|{\mathcal A}|^4|\beta_{\d\u}|^2|\beta_{\u\d}|^2|\u_p,\d_q\re\le\u_p,\d_q|\nonumber\\
&+|{\mathcal A}|^2|\beta_{\d\u}|^2\left(|\beta_{\u\d}|^4+1\right) |\u_p,\u\d_q\re\le\u_p,\u\d_q|\nonumber\\
&+|{\mathcal A}|^6|\beta_{\d\u}|^2|\d_p,0_q\re\le\d_p,0_q|
+2{{\mathcal A}^*}^3{\mathcal A}|\beta_{\d\u}|^2|\d_p,0_q\re\le\u\d_p,\u_q|
\nonumber\\
&+|{\mathcal A}|^4|\beta_{\d\u}|^2|\beta_{\u\d}|^2|\d_p,\u_q\re\le\d_p,\u_q|\nonumber\\
&+|{\mathcal A}|^4 \left(|\beta_{\d\u}|^4+1\right) |\d_p,\d_q\re\le\d_p,\d_q|
+2{{\mathcal A}^*}^2 \left(|\beta_{\d\u}|^4+1\right) |\d_p,\d_q\re\le\u\d_p,\u\d_q| \nonumber\\
&+|{\mathcal A}|^2|\beta_{\d\u}|^2 \left(|\beta_{\u\d}|^4+1\right) |\d_p,\u\d_q\re\le\d_p,\u\d_q| \nonumber \\
&+|{\mathcal A}|^4|\beta_{\d\u}|^2|\beta_{\u\d}|^2|\u\d_p,0_q\re\le\u\d_p,0_q|\nonumber\\
&+|{\mathcal A}|^2|\beta_{\d\u}|^2 \left(|\beta_{\u\d}|^4+1\right) |\u\d_p,\u_q\re\le\u\d_p,\u_q| \nonumber  \\
&+|{\mathcal A}|^2|\beta_{\d\u}|^2 \left(|\beta_{\u\d}|^4+1\right) |\u\d_p,\d_q\re\le\u\d_p,\d_q|\nonumber\\
&+\left(|\beta_{\d\u}|^4+1\right)\left(|\beta_{\u\d}|^4+1\right) |\u\d_p,\u\d_q\re\le\u\d_p,\u\d_q| \Big] +{\rm h.c.},
\label{rhooutpq}
\end{align}
living in the Hilbert space $\mathbb{C}^4\times \mathbb{C}^4$.
Taking the basis $\{|0_p\re, |\u_p\re, |\d_p\re, |\u\d_p\re\}\otimes \{|0_{q}\re, |\u_{q}\re, |\d_{q}\re, |\u\d_{q}\re\}$
we can represent it  as a $16\times 16$ matrix.
Then evaluating the eigenvalues of its partially transposed,
according to \eqref{deflogneg} and
accounting for \eqref{An},
we arrive at the logarithmic negativity in terms of $n$
\b
\label{LNspindegrad}
L_N\left(\varrho^{(out)}_{p,q}\right)=\left\{
\begin{array}{lcc}
2\log_2\left[\frac{32-8n+n^2}{16}\right] & {\rm{if}} & 0 \leq n \leq 4(\sqrt{2}-1)
\\
\\
\log_2\left[\frac{1792-768 n+192 n^2-32 n^3+3 n^4}{512}\right] & {\rm{if}}  &  4(\sqrt{2}-1) \leq n \leq 4
\end{array} \right. \,.
\e
Notice that $L_N$ is a monotonically decreasing function of $n$ starting from maximum value (two) at $n=0$ and reaching the  minimum value (zero) for $n=4$. Although it is continuous, it presents a kink at $n=4(\sqrt{2}-1)$ where
its first derivative is discontinuous.  This is due to the fact that some of the eigenvalues of $\left(\varrho^{(out)}_{pq}\right)^{PT}$ changes the sign at $n=4(\sqrt{2}-1)$.


\subsection{Entanglement degradation for spin-less particles}

For spin-less particles the maximally entangled state corresponding to \eqref{maxentpq} reads
\b
|\Phi\re_{in}=\frac{1}{\sqrt{2}}\left(|0_p,0_q\re+|1_p,1_q\re\right)_{in}|0_{-p},0_{-q}\re_{in}.
\label{PhiAB}
\e
To describe what $out$ observer  sees we use \eqref{invac} and \eqref{in1p} on each mode
and obtain
\ba
|\Phi\re_{in}&=&\frac{1}{\sqrt{2}(1+|\beta/{\cal A}|^2)}\left(|0_p,0_q;0_{-p},0_{-q}\re
-\frac{\beta^*}{\cal A^*} |0_p,1_q;0_{-p},1_{-q}\re
-\frac{\beta^*}{\cal A^*}|1_p,0_q;1_{-p},0_{-q}\re+\frac{\beta^{*2}}{{\cal A}^{*2}} |1_p,1_q;1_{-p},1_{-q}\re\right)_{out}\notag\\
&+&\frac{1}{\sqrt{2}}\left( |1_p,1_q;0_{-p},0_{-q}\re \right)_{out},
\label{Phipp}
\ea
Also in this case we have assumed $\beta_{d,-d}(p )\approx\beta_{d,-d}(q)$ and ${\mathcal{A}}(p )\approx {\mathcal{A}}(q)$. Then, the density operator for particles modes $pq$,
can be derived by tracing $|\Phi\re_{in}\le\Phi|$
over anti-particle modes with momentum $-p$ and $-q$ in the $out$ region $\varrho^{(out)}_{p,q}={\rm Tr}_{-p-q} (|\Phi\re_{in}\le \Phi|)$.  It results
\begin{align}
\varrho^{(out)}_{p,q}=\frac{1}{4}\Big[&
\frac{|{\cal A}|^4}{( |{\cal A}|^2+|\beta|^2)^2} |0_p,0_q\re\le 0_p,0_q|
+\frac{2 |{\cal A}|^2}{|{\cal A}|^2+|\beta|^2}
|0_p,0_q\re\le 1_p,1_q|
+\frac{|{\cal A}|^2 |\beta|^2}{( |{\cal A}|^2+|\beta|^2)^2}
|0_p,1_q\re\le 0_p,1_q| \nonumber\\
&+\frac{|{\cal A}|^2 |\beta|^2}{( |{\cal A}|^2+|\beta|^2)^2}
|1_p,0_q\re\le 1_p,0_q|
+ +\frac{2 ( |{\cal A}|^2+|\beta|^2 )^2 |\beta|^4}{ 2 ( |{\cal A}|^2+|\beta|^2 )^2}
|1_p,1_q\re\le 1_p,1_q| \Big] + {\rm h.c.},
\label{rhoAB}
\end{align}
living on the Hilbert space $\mathbb{C}^2\otimes\mathbb{C}^2$.

Proceeding like in Section \ref{entaC} we arrive at
\b
L_N(\rho^{out})=\log_2\left[1+\frac{(2-n)^2}{4}\right].
\label{Nn}
\e
In this case $L_N$ is a smoothly decreasing function of $n$ starting from maximum value (one) at $n=0$ and reaching the minimum value (zero) for $n=2$.


\section{Conclusion}\label{conclu}

In summary we have addressed the issue of entanglement transformations in an expanding spacetime characterized by Robertson-Walker metric. Being the latter flat in the remote past and far future we have first provided Bogolyubov transformations for spin-particles between these two asymptotic regions.
We have then shown particles-antiparticles (modes $p$ and $-p$) entanglement creation when passing from remote past to far future as well as particles (modes $p$ and $q$) entanglement degradation.
For such cases we have derived analytical expressions of logarithmic negativity (for spin particles as well as for spin-less ones) as function of the density $n$ of created particles, thus relating entanglement transformations to particles creation phenomenon.

It turns out that in the presence of spin the amount of created entanglement behaves qualitatively the same of spin-less case.
Quantitative differences arise to different dimensionality of the involved bipartite spaces (since $\max L_N=\log_2(d^2)$ for bipartite systems living in $\mathbb{C}^d\otimes \mathbb{C}^d$) and to different relations of Bogolyubov coefficients with the density of created particles (see \eqref{An} and \eqref{An2}).

Conversely for entanglement degradation there are also qualitative differences between spin and spin-less particles cases.
As matter of fact the decreasing rate of entanglement of a remote maximally entangled state of spin particles presents a discontinuity.
This should be ascribed to the presence of particles in remote past as well as to spin degrees of freedom. These two ingredients increase the number of entries in
 the matrix representing the density operator and due to their specific dependence from $n$ it happens that some eigenvalues of its partially transposed change the sign while varying $n$.

In conclusion, we have given evidences that entanglement transformations are inherently related to the spacetime structure and highlighted the role of spin degrees of freedom.
The found results besides deepening our understanding about entanglement,
could also offer the perspective of using entanglement as a tool to learn about curved spacetime features.
Applications to specific cosmological models with the goal of deducing
cosmological parameters from entanglement remain matter for future studies.


\section{Acknowledgments}\label{Acknow}

S. Moradi would like to thank the University of Camerino for kind hospitality and INFN for financial support.


\newpage

\end{document}